\documentclass[aps,prl,twocolumn,10 pt,showpacs]{revtex4}
\usepackage{amssymb}
\usepackage{graphicx}

\begin{document}

\title{Longitudinal and transverse noise in a moving Vortex Lattice}
\author{J. Scola, A. Pautrat, C. Goupil, Ch. Simon}
\affiliation{CRISMAT/ENSI-Caen, UMR 6508 du CNRS,6 Bd Marechal Juin, 14050 Caen, France.}

\begin{abstract}
We have studied the longitudinal and the transverse velocity fluctuations of a moving
vortex lattice (VL) driven by a transport current. They exhibit both the same broad
spectrum and the same order of magnitude. These two components are insensitive to the
velocity and to a small bulk perturbation. This means that no bulk averaging over the
disorder and no VL crystallization are observed. This is consistently explained referring
to a previously proposed noisy flow of surface current whose elementary fluctuator is
measured isotropic.
\end{abstract}


\newpage
\maketitle

Recent theoretical studies have pointed out that the VL, as an example of driven
disordered system, could exhibit different topological order during its motion
\cite{vortexphases}. The experimental problem is to have access to a signature of
disorder in the VL. Voltage and magnetic field fluctuations measurements appear to be a
quite natural tool. Indeed, it has been established for years that such noise
measurements turned out to be very efficient to collect information on the dynamical
behavior of a moving Vortex Lattice (VL) and on the way it can be associated to its
pinning properties \cite{clem}. In order to analyze the fluctuating part of the VL
submitted to the driving force, the most direct experiment consists in a measure of the
noisy electro-magnetic fields for different points of a voltage versus current (V(I))
characteristic. The dissipative part of this curve usually presents two regimes. Just
above the depinning threshold (critical current $I_{c}$), in the Low Current Regime
(LCR), the average voltage response does not scale as $(I-I_{c})$. This implies
inhomogeneous depinning, i.e. different onset of motion, coming either from intrinsic
reasons (''plastic phase'')\cite{shobo}, from extrinsic reasons (simple dispersion of
critical current)\cite{sans}, or eventually from both of them. Nevertheless, in each case
this can be formalized as a plastic-like flow with VL chunks moving at different
velocities. When increasing the current, the linear regime (the $\emph{flux-flow}$ where
dV/dI is a constant) is reached. This $\emph{flux-flow}$ regime corresponds to the whole
VL in motion. Its long time averaged movement is coherent, which justifies the
description in terms of an elastic response of an ordered media. If one supposes that the
pinned state is disordered, one can realize that the VL should average its pinning
efficiency through disorder to finally order at a threshold current. This crossover
between two dynamical states can be formalized in terms of a dynamical crystallization
\cite{koshelev}. Including the periodicity of the VL, the formation of elastic channels
with transverse barriers at high velocities is predicted \cite {BrG}. Numerical
simulations support this picture of flowing channels \cite {olson} and some give rise to
a growth of the transverse order driven by the current, leading to a transverse freezing
\cite{kolton}. A common point between those predictions is that a high drive implies a
healing of defects present in the VL. This is expressed in a dynamical averaging at least
in one of the direction in the plane of the flow. A loss of noise, i.e. a loss of
interaction with the pinning centers, is thus expected. We note that in contrast with the
very active field of theoretical work and numerical simulations, only very few
experiments have been devoted to the verifications of the preceding points. There are
even opposite results. For example, Pla\c{c}ais $\textit{et al}$ investigated the high
current regime without finding any decrease of the voltage longitudinal noise
\cite{placais}. In view of the above cited theories, it appears that the fluctuations in
the direction perpendicular to the flow should also contain pertinent signatures. Maeda
$\textit{et al}$ measured the correlation of the fluctuations in the flux density along
this direction \cite{maeda}, but did not extend into analyzing the transverse noise
itself when the driving force is increased.

More precisely, most of the interest resides in the VL correlation functions for the
theories or in the positions of vortices for the simulations. The associated fluctuating
quantities are essentially the velocities. As the vortex flow is associated to
dissipation, theoretical predictions can be checked by measuring the voltage noise. A
first important issue is to collect what really corresponds to velocity fluctuations when
measuring the voltage noise along a path that connects the two voltage contacts. Indeed,
irrespective of any precise noise model, a look at the Josephson equation $%
\mathbf{E}=-\mathbf{v_{L}}\wedge \mathbf{B}$ evidences that both velocity fluctuations
$\delta \mathbf{v_{L}}$ and magnetic field fluctuations $\delta \mathbf{B}$ can play a
role. This discrimination between $\delta \mathbf{v_{L}}$ and $\delta \mathbf{B}$ has
been a central point for the understanding of the origin of the VL noise. Historically,
first experiments which gave evidence that the voltage noise was coming from vortex
motion were performed by Van Ooijen and Van Gurp \cite{vangurp}. They interpreted their
results as shot-noise implying strong $\delta \mathbf{B}$. The central idea is that flux
bundles with short range correlation are generating pulse voltages with finite lifetime.
It could have clearly demonstrate the existence of flux bundles, but in spite of numerous
developments \cite{gray}, this model has not been confirmed by experiments
\cite{clem,placais,weissman}. Discriminating tests which invalidate this ''flux bundle''
approach are the absence of correlation between magnetic field noise and voltage noise in
the $\textit{ flux-flow}$ regime and the smallness of the magnetic field noise \cite
{heiden}, whereas the shot noise analogy predicts strong correlations and large field
noise \cite{placais}. This leads to the conclusion that the moving VL noise is not
generated through local density fluctuations, which is not consistent with flux-bundles
as independent entities \cite{placais}. If the magnetic field noise is not at the origin
of the moving VL noise, the other scenario is pure vortex velocity fluctuations $\delta
\mathbf{v_{L}} $ \cite{heiden}. Unfortunately, the simple picture of a quasi-perfect 2D
moving lattice cannot describe the field noise, and consequently cannot explain the
absence of correlation between field noise and voltage noise. In order to answer to this
latter question, it is necessary to know and to locate the fluctuators.
Cross-correlations experiments in the $\textit{flux-flow}$ regime in low T$_{c}$ alloys
and metal strongly suggest that there are surface current fluctuations \cite{placais}. Now
in the region close to the peak effect, additional large voltage fluctuations are present
and are associated with non Gaussian averaging of the noise \cite{marley, rabin}. A model
of this excess noise proposes a dynamical mixture of two VL phases \cite {paltiel}.

Our present study deals with a more conventional case, i.e. the study of the VL noise
when a unique VL phase is present.
First we propose to isolate the velocity fluctuations
with a special care to the component perpendicular to the direction of the flow.
As far as we know, the response of this component to the driving force has never been
experimentally investigated and compared to the predictions.
To fulfill this gap would bring precious hints on how the vortex order is determined by the velocity.
In particular, we show that the fluctuations stand without averaging, meaning that no cristallization is observed. Furthermore, this noise regime is not affected by an artificial bulk perturbation, but turns out to be dominated by surface effects.

\section{Sample and experimental resolution}

All data presented here are measured using a sample of Pb-In (10.5\% of In by weight,
size $12.4\times 4.1 \times 0.15mm^3$). All basic parameters are in agreement with
tabulated values ($\rho (T_c)=6.15 \mu\Omega .cm$, $T_{c}=7K$, $B_{c2}(4.2K)=0.29T$)
\cite{farrel}. This ensures the good bulk homogeneity of the sample. As usual for a
metallic alloy, the sample exhibits a mirror-like shape at the optical scale and AFM
inspection evidences a moderate surface roughness at the scale $0.1-1 \mu m$ (mostly self
similar surface with a corrugation of about 10 nm over 100 nm in this scale). Our
experimental set-up is drawn in figure 1. The space between the longitudinal and
transverse contacts is respectively d= 4mm and 1 mm. The sample was supplied by noise
free current made by car batteries and thermalized power-resistances. Noisy Voltages were
recorded and amplified by ultra low noise preamplifier
(SA-400F3) with a resolution of $0.7 nv/ \sqrt{Hz}$. Magnetic flux noise $%
\delta B_z$ was picked-up by a 10 turns-coil largely surrounding the sample, so as to
avoid a non perfect coupling \cite{placais}. The signal was then amplified by an original
set up consisting in a highly linear transformer (Vitrovac) with turns ratio
($1/1000$), coupled with a low current noise amplifier (INA114). Taking care of external
electromagnetic
perturbations, it was possible to measure field fluctuations less than one $%
\mu G / \sqrt{Hz}$.

\subsection{Velocity noise measurement procedure}

\subsubsection{Numerical representation}

The analog signals $u_i(t)$ at the input of the acquisition card of the
computer are converted into digital signals and then numerically processed.
Since vortex noise is a random signal, power spectra are not relevant and
one must consider the autocorrelation function of the noise instead:
\begin{equation}
A_{ii}(\tau) =\lim_{T \rightarrow \infty} \int_{-T/2}^{T/2} u_i(t) u_i(t+\tau) dt.
\label{autocorrelation}
\end{equation}

According to the Wiener-Kintchine theorem, the Fourier Transform of the
autocorrelation function is the Power Spectral Density (PSD) :
\begin{equation}
S_{ii}(f)= \int_{-\infty}^{\infty}A_{ii}(t)e^{-2j\pi ft}dt \equiv \lim_{T \rightarrow
\infty} \langle \frac{ U_i(f) U^*_i(f) } {T} \rangle. \label{psd}
\end{equation}

We did not focus on the shape of the power density spectra because it does
not vary much with magnetic field or current in our experimental conditions.
In this paper, we represent noise either by the PSD ($\delta U(f)$) or by
the PSD integrated over the frequency bandwidth ($\delta U* =
\int_{10}^{1000} \delta U(f) df$), which corresponds to the rms noise value.
The detail of the spectra envelop will be discussed in later works.

\subsubsection{The Josephson equation}

\label{josephsonequation} As stated in the introduction, as we are interested in velocity
fluctuations, it is necessary to isolate the different noisy fields. In our experiments,
we measure physical quantities averaged over large lengthscales (the sample is relatively
large and the distance between the voltage pads is about few millimeters). Such mean
quantities are properly described by the Josephson equation:

\begin{equation}
\mathbf{{E} = -{v_L} \times {B}.}  \label{josephson}
\end{equation}

In our geometry (see Fig.\ref{setup}), equation \ref{josephson} can be
differentiated as follows :
\begin{equation}
\delta E_{long} = \delta E_y = \delta v_{Lx} B_z + v_{Lx} \delta B_z
\label{jojolong}
\end{equation}
\begin{equation}
\delta E_{trans} = \delta E_x = \delta v_{Ly} B_z + v_{Ly} \delta B_z
\label{jojotrans}
\end{equation}

In \emph{flux-flow}, the Hall voltage is negligibly small so that $v_{Ly} \approx 0$, and
the mean electric field can be written :
\begin{equation}
\langle E \rangle \approx E_{long} = v_{Lx} B_z = R_{FF}(I-I_c)/d,
\label{vlong}
\end{equation}

where $R_{FF}$ stands for the \emph{flux-flow} resistance, and $d$ is the
distance between the voltage pads.

Putting \ref{vlong} into \ref{jojolong}, one obtains
\begin{equation}
\delta E_{long} = \delta v_{Lx}B_z + \frac{R_{FF}(I-I_c)}{B_z d}\delta B_z
\label{Elong}
\end{equation}
\begin{equation}
\delta E_{trans} = \delta v_{Ly} B_z.  \label{Etrans}
\end{equation}

This relation between the voltage noise and the velocity fluctuations is $%
\textit{a priori}$ valid for any noise model, simply assuming that Josephson relation
applies at our experimental length scale (millimeter
scale). This does not depend on the source of $\delta \mathbf{v_L}$ versus $%
\delta \mathbf{B}$. Looking at the equation (8), one can realize that the
transverse voltage noise gives a direct measurement of the velocity
fluctuations in the $y$ direction. Yet, the longitudinal voltage noise has
an extra contribution involving the magnetic field fluctuations. In order to
collect the velocity fluctuations in the $x$ direction, one should measure
simultaneously the longitudinal voltage noise and the magnetic field noise
and then subtract the magnetic field component.

\section{Experimental Results}

In this study, we report on the influence of the driving force on the in-plane
fluctuations of the VL velocity ($\delta v_{Lx}$ and $\delta v_{Ly}$) at 4.2 K, for
several magnetic fields. To begin with, we check the experimental validity of the
above-described procedure. The study is then divided into two parts. In a first part, we
present experiments with a DC driving force : the velocity fluctuations are measured for
different points of the I-V characteristics. Secondly, we discuss the noisy response of a
moving lattice driven by a small perturbation force.

\subsection{Velocity fluctuation measurement}

As a preliminary step, we check the experimental validity of the equations (7) and (8).
The longitudinal electric field rms noise ($\delta E^{\ast }$) is collected for different
currents in the \emph{flux-flow} regime, and reported in the figure \ref{procedure2300}a.
The results are similar if the noise is considered at a given frequency rather than
integrated over the whole frequency bandwidth. In order to determine the \emph{flux flow}
regime where the Josephson equation applies at the measurement scale, the V(I) curve is
also drawn in the figure \ref{procedure2300}b. This corresponds to the regime where the
differential resistance is a constant. It can be realized from the experimental data that
$\delta E^{\ast }$ can be divided into two terms: a constant term and a term which varies
linearly with ($I-I_{c}$). This observation stands for all the magnetic fields we have
investigated (from $0.32H_{c2}$ to $0.93H_{c2}$). An identification with the Josephson
equation (7) suggests that the two fluctuating components $\delta v_{Lx}$ and $\delta
B_{z}$ are constant with respect to ($I-I_{c}$). This result is confirmed by a direct
measurement of the magnetic field noise $\delta B_{z}$ by the pick-up coil. The obtained
value is then compared to
the estimation of $\delta B_{z}$ calculated from the slope of the $\delta E_{y}$ Vs. ($%
I-I_{c}$) curve, using (7). The agreement was very satisfactory for all the magnetic
fields at which we made measurements; the result reported in the figure
\ref{procedure2300}a corresponds for example to $0.23T=0.7H_{c2}$. In the rest of the
paper, $\delta B_{z}$ will refer either to the directly measured value or to the
estimation from the slope. On the other hand, the transverse component of the noisy
electric field is measured constant in $\textit{flux-flow}$, as predicted by the
equations (8). This shows an interesting property: a measure of pure velocity
fluctuations. It will be analyzed in more details below.

\subsection{Noise in DC biasing}

Figures \ref{EIplus}a and \ref{EIplus}b show the detailed results in both directions for
different DC currents ($B=0.1T$). The fluctuations appear at the first dissipative
current, i.e when the VL starts to move. In the non linear part of the $E(I)$ curve (fig.
3c), the longitudinal fluctuations exhibit a fuzzy behavior. In this range of driving
forces the whole VL is not in \emph{flux-flow} yet, and the Josephson equation is not
valid at the sample scale. Neutron experiments have pointed out that inhomogeneity of the
critical current can lead to the following depinning \cite{sans}: slices of VL along
which the critical conditions are similar depin in sequence, until the whole VL is in
\emph{flux-flow}. Therefore, the longitudinal noise signature in this range of currents
can be seen as a succession of depinning peaks. This mimics plastic deformations such as
those observed through fingerprints in the differential resistance in $NbSe_{2}$
\cite{vortexphases}. The fact that the longitudinal noise exhibits a more jagged behavior
than the transverse one can be explained by an excess of magnetic field noise due to
fluctuations in the number of (moving) vortices. The velocity, even if spatially
inhomogeneous, would not fluctuate much more than in $\textit{flux-flow}$.

As soon as the $\textit{flux-flow}$ is reached, equations (7) and (8) apply, and $\delta
v_{Lx}$ and $\delta v_{Ly}$ can be extracted from the electric field noise (fig.
\ref{deltaVLdevit}). We observe that the velocity fluctuations in the longitudinal
direction do not depend on the current. $\delta v_{Ly}$ reveals the same behavior in the
transverse direction. In addition the ratio $\alpha =\frac{\delta v_{Ly}}{\delta v_{Lx}}$
is constant and equals $0.5\pm 0.1$. This means that the velocity fluctuations are large
in the two directions. More importantly, they are not averaged by the motion. It must be
emphasized that neutron scattering experiments carried out in a similar sample give the
evidence of a well ordered VL (crystalline like) in the same conditions \cite{houston}.
Thus we study the noise signature of a moving crystal of vortices or a moving Bragg Glass
of vortices, i.e. the high velocity ordered state seen in simulations. Nevertheless, it
is important to realize that large velocity fluctuations in the two directions (both
longitudinal and transverse to the motion) are present in this regime. The velocity
independence of these fluctuations shows that the disorder responsible for these velocity
fluctuations is not averaged to zero. This contrasts with the disappearance of the
fluctuating part of the pinning component as predicted in the dynamic crystallization
developed in \cite{koshelev}. For a VL propagation through channels at high driving
force, large transverse barriers are expected to keep the channels rigid. If the large
transverse noise in the LCR with $\alpha >1$ is in a qualitative agreement with the
simulations of Kolton $\textit{et al.}$, no transverse velocity fluctuations are expected
in flux-flow ($\emph{transverse freezing}$) whereas we observe substantial ones. The
persistence of an equivalent transverse noise power over the whole range of current,
outside the depinning peaks (fig. \ref{EIplus}), tends to prove that the nature of flow
is not fundamentally different in the LCR and in FF. We conclude that the measured noise
signatures are not consistent with a dynamically induced phenomena with a healing of
defects in the VL. This has to be brought close to the simple fact that the mean DC
response of the sample is strongly non-ohmic and the critical current does not disappear
at high drive. The system keeps the memory of its pinned configuration, i.e the pinning
force does not disappear with the increase of the velocity. Even in motion, the VL still
interacts with the pinning sites: the critical current remains and noise is generated. As
a result, vortex noise can be fundamentally decomposed into a static part (the ''memory''
of the system) and possibly a dynamical part which expresses the dependence of the
interactions on the mean velocity of the lattice. But as no velocity dependence is
observed here, both velocity fluctuation components originate from fluctuations of the
pinning force which are not influenced by the mean velocity of the lattice.

\subsection{Static noise Vs. dynamical noise}

The question of the origin of both transverse and velocity fluctuations are thus linked
to the very nature of the pinning. We recall that in Pb-In the pinning properties are
dominated by the (quite standard) surface roughness \cite{sans}. A consistent noise model
has been proposed and the surface origin of the fluctuations evidenced \cite{placais}.
While in motion, the VL experiments the roughness of the surface and consequently, the
boundary conditions are modified in time and space. The VL explores randomly the
different metastable pinning configurations, and the critical current (or surface
current) fluctuates locally and temporarily, in absolute value and in direction. Such
surface current fluctuations are compensated by opposite bulk current fluctuations in
order to keep constant the total transport current inside the sample. Velocity
fluctuations are then generated along with the noisy component of the driving force.
Besides, the noisy bulk current induces a [possibly substantial] magnetic field noise on
behalf of Maxwell law. The surface current fluctuations behave
like a noise generator of vortex velocity and density. As a consequence, $%
\delta v_{Lx,y}$ and $\delta B_{z}$ have the same spectra \cite{placais}.
This prediction is verified in our sample (figure \ref{idemspectres}).

From a quantitative point of view, it is also predicted that the amount of noise is
determined by the correlation length $C$ of the surface supercurrent. More precisely,
with $\delta V_{L}=R_{FF}\delta I_{c}/d.B$, and in the simplest case of 2D homogeneous
and stationary fluctuations, one can write $\delta I_{c}\approx I_{c}\sqrt{C_{x}C_{y}/S}$
with $S$ the surface of the sample limited by the voltage pads, and $C_{x,y}$ the
correlation length. $C_{x,y}$ is the unique adjustable parameter. We verified the
stability of the fluctuations by measuring the second-order spectrum $S^{(2)}(f_{2})$,
the spectrum of noise spectra \cite{note}. The voltage signal was acquired during a very
long time (about an hour) then segmented, and finally each segment was Fourier
transformed. Time series of noise power were taken for different ranges of frequencies (a
few Hertz wide), and Fourier transformed over a $2mHz-1Hz$ spectral bandwidth. We observe
essentially a white spectral density, confirming the stability of the process. We
obtain $\sqrt{C_{x}C_{y}}\approx 4-0.5\mu m$ for applied field ranging from $0.23H_{c2}$
to $0.93H_{c2}$. This range of values is realistic since it lies between the inter-vortex
distance and the sample size. The order of magnitude of the size of the correlation
length is also in very good agreement with the values found in \cite{placais} at lower
temperature. As we measure here the two components of the velocity fluctuations, one have
access to the vectorial form of the fluctuators. With a two dimensional form for the
spatial correlation length
and using the experimental result $\frac{\delta v_{Ly}}{\delta v_{Lx}}%
=0.5\pm 0.1$, one find that $C_{x}= 1 (\pm 0.3)C_{y}$. The correlated domain of the
surface current is finally found isotropic, what fits well with the idea that the surface
is randomly explored and offers equivalent boundary conditions in all direction.

It appears that the transverse and low frequency broad band noise (BBN) can be understood
as a part of a global noise mechanism driven by a noisy surface current. It remains that
the bulk of the sample is obviously not
free from defects. As soon as the current penetrates the bulk, i.e. for $%
I>I_{c}$, the VL flow can interact with bulk defects. The reason why dynamically induced
phenomena such as disorder averaged by the velocity, typical of a bulk process, are not
observed has to be discussed. One can propose that the surface driven noise intensity
strongly dominates a possible bulk driven noise, or that bulk signatures are at much
higher frequencies (about MHz for Washboard-like signature under similar experimental
situations). To go deeper inside this question, one can superimpose low frequency bulk
perturbations in order to see if the noise is influenced. This experimental configuration
originally comes from a technical hitch. Car batteries and connections turned out to
require long thermalization time before being completely noise free. Otherwise, one
observe an excess of current noise $\Delta$I in the longitudinal spectrum, which is
simply due to the linear superposition of this noisy supply current $\Delta$I on the
noisy current due to the vortices. It is striking to realize that no trace of this
spurious noise is observed in the transverse spectrum. These experiment shows that
superimposing a noisy lorentz force to the motion does not change the underlying velocity
fluctuations. Furthermore, we applied a controlled sinusoidal force to the VL in
\emph{flux-flow}. The AC current applied is denoted $i_{ac}sin(2\pi f_{d}t)$ with
$i_{ac}$ such as $v_{ac}=R_{FF}i_{ac}$ is of the order of magnitude of the voltage noise
at the frequency $f_{d}$.  Low frequency values ($f<2017Hz$) are employed to avoid skin
effect in order to be sure to perturb the bulk of the sample. The figure \ref{compobruy}
represents an example of the longitudinal and transverse velocity spectra with and
without a low frequency sinusoidal component. For $i_{ac}>0$, all the low frequency BBN
is preserved and $v_{ac}$ is \emph{entirely} dissipated in the $y$ direction. The AC
contribution to the velocity fluctuations simply stacks linearly to the noise regime but
the broad band spectra are non sensitive to this bulk perturbation. The bulk response of
the sample is thus decoupled from the ''static'' and apparently robust noise regime.
Compared to the surface, the bulk seems to be a quiet host for the VL as far as low
frequency BBN in $\textit{flux-flow}$ is involved.

In conclusion, we have investigated the longitudinal and transverse components of the
electric field noise generated by a moving vortex lattice in a low $T_{c}$ sample. The
transverse component was shown to contain only velocity fluctuations, whereas the
longitudinal one contains also the noisy magnetic field contribution and depends on the
mean vortex velocity. The velocity fluctuations do not show any averaging effect in both
directions when increasing the lattice velocity far inside the $\textit{flux-flow}$
regime. In addition, they are not affected by a noisy bulk force or by a small AC bulk
perturbation. This agrees with fluctuations originating from the surface, and shows the
small sensitivity of these fluctuations to bulk perturbations. A quantitative analysis
provides a picture of isotropic noisy superficial current, in agreement with the model
proposed in \cite{placais}. We notice also that the study of the fluctuations
perpendicular to the motion seems to be particularly appropriate to probe the intrinsic
fluctuations sources in superconductors (and possibly other dynamical systems).

Acknowledgments: We wish to acknowledge Bernard Pla\c{c}ais (ENS Paris, France) for his
advises during the adjustment of this experiment. This experiment was supported by "la
r$\acute{e}$gion basse Normandie".

%
{\small

\newpage
\begin{figure}[tbp]
\centering
\includegraphics*[width=6cm]{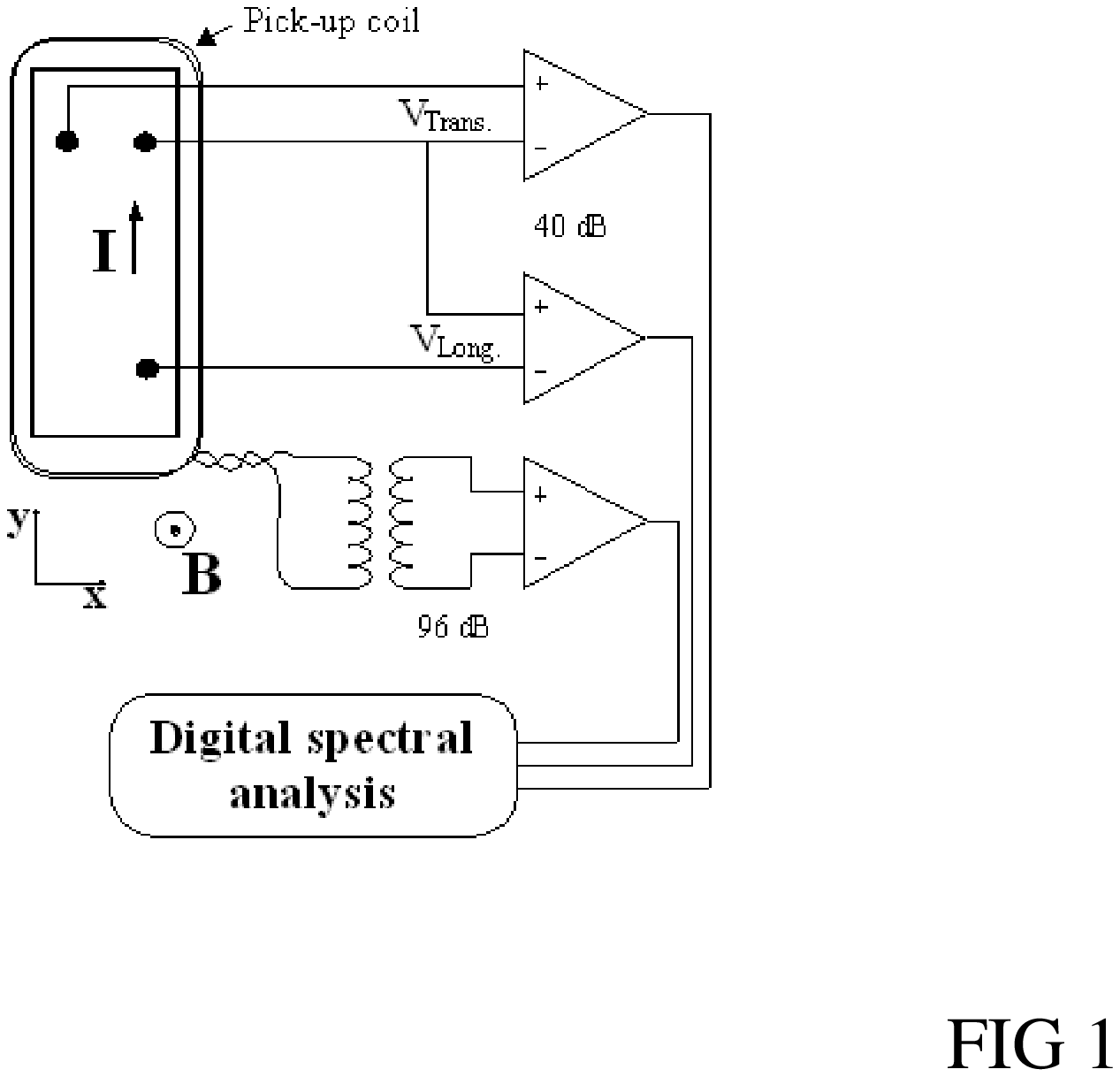}
\caption{Electric field and magnetic field noise experimental set-up. The current is
supplied by batteries and yields noise-free currents. All amplifying equipment are
electromagnetically shielded.} \label{setup}
\end{figure}

\begin{figure}[tbp]
\begin{center}
\includegraphics*[angle=0,width=6cm]{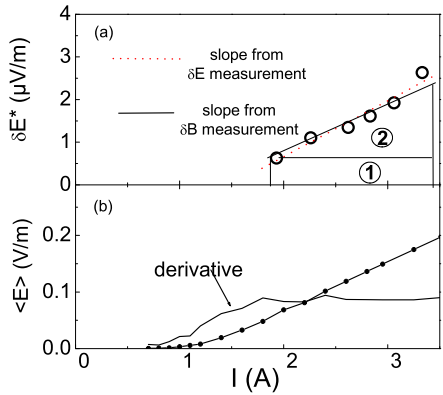}
\end{center}
\caption{(a) Electric field noise integrated over two frequency decades ($%
10-1000 Hz$) $\protect\delta E^*$, plotted against the current ($\circ $) (4.2K, 0.23T).
The dashed line represents the linear fit of the FF noise, yielding $\protect%
\delta B^*_z = 32 mG$ using Eq. (7). The solid line represents the slope calculated from
the direct $\protect\delta B_z$ measurement. The background noise integrated over two
decades has been subtracted. (1) represents the amount of excess noise independent of the
current and (2) the amount of noise dependent of the current as explained in the text.
(b) Solid line and $\circ $ : mean electric field $\langle E_y \rangle$ against the
current. The thin line represents $dE/dI$ against the current. The main critical current
is defined by the extrapolation of the linear part of the E(I) curve.}
\label{procedure2300}
\end{figure}

\begin{figure}[tbp]
\begin{center}
\includegraphics*[angle=0,width=6cm]{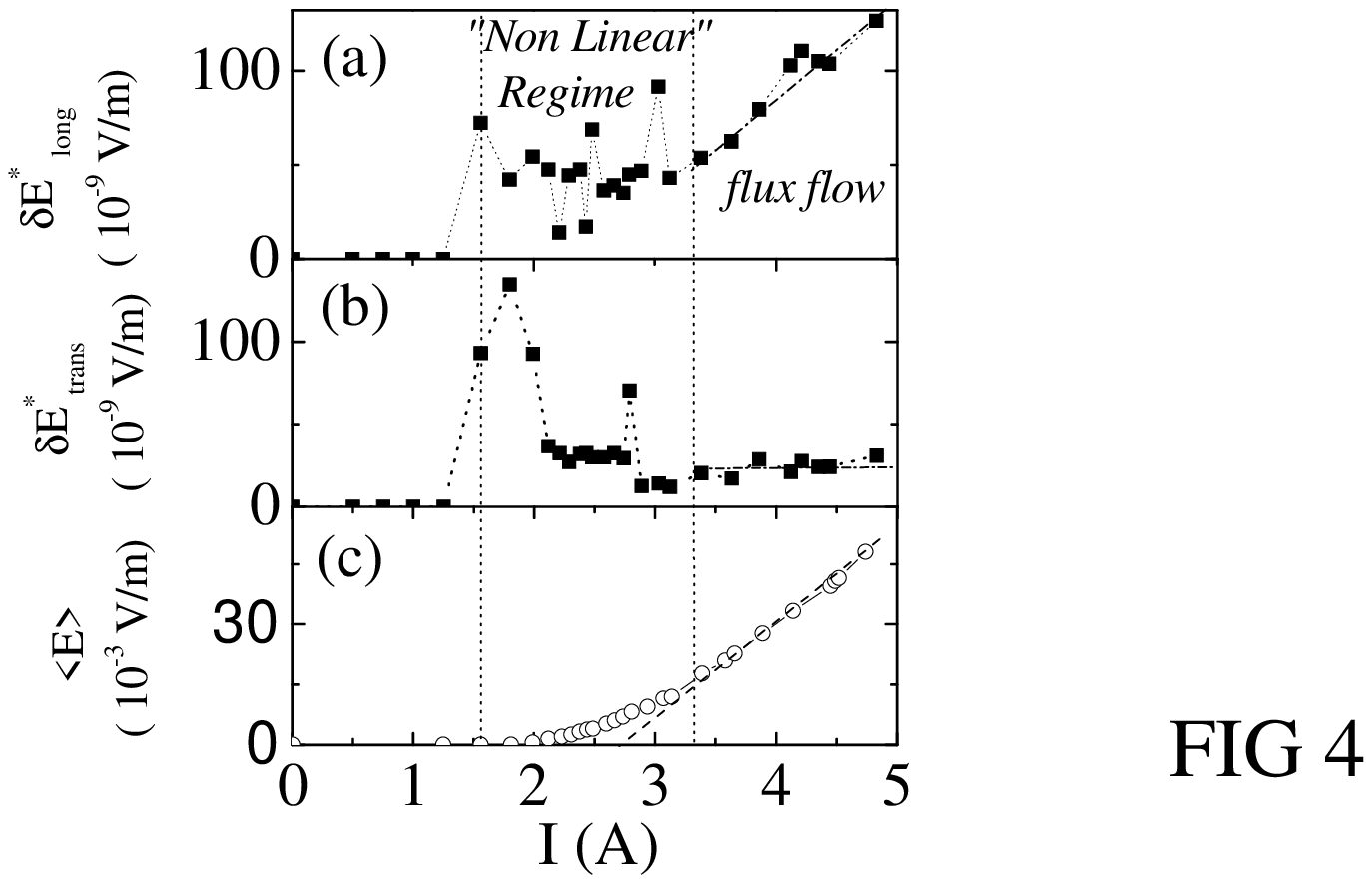}
\end{center}
\caption{Electric field noise power spectral density integrated over two decades
($10-1000Hz$) and plotted against the current for the different dynamical regimes
(T=4.2K,B=0.1T) : (a) in the $y$ direction ($E*_{long}$) and (b) in the $x$ direction
($E*_{trans}$). Dashed lines are guides for the eyes. The mean electric field $\langle E
\rangle$ is represented against the current in (c). $\langle E \rangle$ is measured in
the $y$ direction; $\langle E_x \rangle= 0$ within our experimental resolution.}
\label{EIplus}
\end{figure}

\begin{figure}[tbp]
\includegraphics[angle=0,width=6cm]{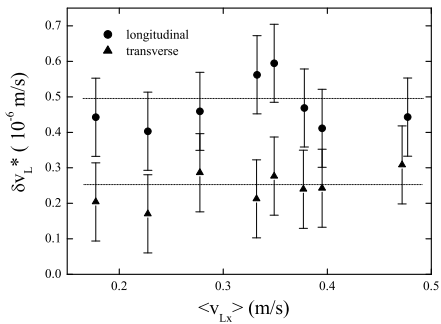}
\caption{velocity fluctuations in the longitudinal ($\bullet$) and transverse
($\blacktriangle$) directions plotted against the mean velocity of the lattice $\langle
v_{Lx} \rangle$ (T=4.2K,B=0.1T). This range of velocities corresponds to the
$\emph{flux-flow}$ regime. Dotted lines are guides for the eyes.} \label{deltaVLdevit}
\end{figure}

\newpage
\begin{figure}[tbp]
\includegraphics[angle=0,width=6cm]{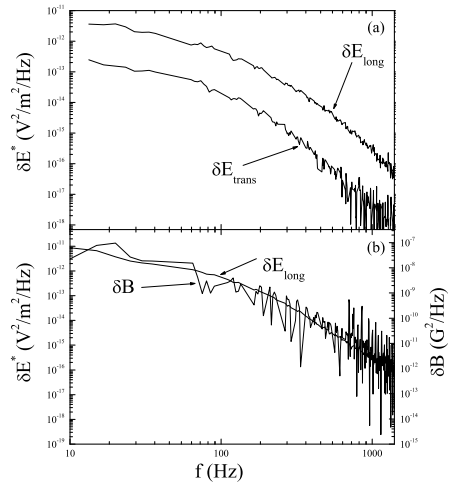}
 \caption{Up: Longitudinal and transverse electric fields noise spectra and down:
Longitudinal electric field noise Magnetic field noise spectra. Both are taken in
$\textit{flux-flow}$ (B= 0.23T, T=4.2 K,I= 6.9A). Note the similarity of the shape for
all the spectra.} \label{idemspectres}
\end{figure}

\begin{figure}[tbp]
\begin{center}
\includegraphics*[angle=0,width=6cm]{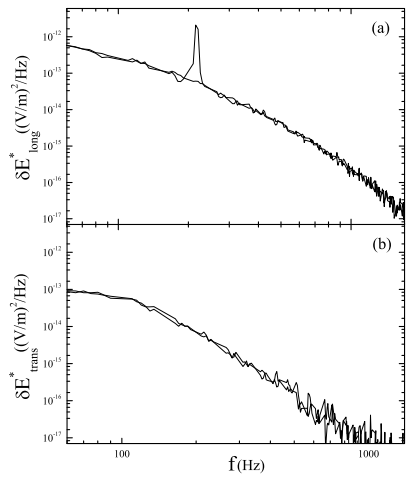}
\end{center}
\caption{ (a) Flux-flow longitudinal noise with and without a superposing ac component.
(b) The corresponding transverse noise, no ac component is observed and the noise is
fully preserved.} \label{compobruy}
\end{figure}

\end{document}